\documentclass[a4paper,dvips,12pt]{article}
\usepackage{epsfig,graphics,graphicx}
\begin{document}
\def\Base{{\cal B}}
\def\Total{{\cal P}}
\def\Hopf{{\cal H}}
\def\Zero{{\bf Z}}
\def\Ji{{\cal J}}
\def\Id{{\bf Id}}
\def\een{{\bf 1}}
\def\dif{{\bf d}}
\def\Complex{{\bf C}}
\def\be{\begin{equation}}
\def\ee{\end{equation}}
\def\eea{\end{eqnarray}}
\def\bea{\begin{eqnarray}}
\def\bean{\begin{eqnarray*}}
\def\eean{\end{eqnarray*}}
\renewcommand{\theequation}{\thesection.\arabic{equation}}
\pagestyle{myheadings}
\markright{\it CO07-1}
\vskip.5in
\begin{center}
\vskip.4in {\Large\bf Classical Principal Fibre Bundles from a Quantum Group Viewpoint}
\vskip.3in
F.J.Vanhecke\footnote{Email: \tt vanhecke@if.ufrj.br}, C.Sigaud\\
Instituto de F\'{\i}sica,UFRJ,Rio de Janeiro\\
A.R.Da Silva\\Instituto de Matem\^{a}tica,UFRJ,Rio de Janeiro
\end{center}
\vskip.2in
\begin{abstract}
In this short article we review\footnote{Communication at the XXIVth Brazilian Meeting on Particles and Fields, Caxamb\'{u}-MG,october 2003.} how the classical theory of principal fibre bundles (PFB) transcribes in an algebraic formalism. In this dual formulation, a PFB is given by a right co-module algebra ${\cal P}$ over a Hopf algebra ${\cal H}$ with a mapping 
$\Delta_R:{\cal P}\rightarrow{\cal P}\otimes{\cal H}$. In our case ${\cal P}$ is the (commutative) ${\bf C}^*$-algebra of complex-valued continuous functions on the {\it total space} $P$ and ${\cal H}$ is the Hopf algebra of complex-valued functions on the {\it structure group} $G$. These underlying spaces are endowed with a topology only. The subalgebra ${\cal B}$ of $\Delta_R$-invariant elements is identified with the algebra of complex-valued functions on the base space $B$. In order to define horizontal one-forms, a differential calculus is needed. Since no a priori differential structure is assumed, we use the calculus of the universal differential envelope $\Omega^\bullet({\cal P})$ which can be defined on any unital algebra. A connection on the PFB is then defined by a splitting of the universal one-forms as a direct sum of horizontal and vertical subspaces : $\Omega^1({\cal P})=\Gamma^1_{hor}\oplus\Gamma^1_{ver}$. In case of a {\it strong connection} in a {\it trivial} PFB, the general expression and gauge transformation of the  connection one-form and the curvature two-form are given. A {\it locally trivial} PFB can be  constructed through a gluing procedure of a cover of the algebra ${\cal P}$ (see this meeting's poster session P112, where examples are given).
\end{abstract}
\section{Topological quantum principal bundles}
\setcounter{equation}{0}
Let $\Total$ be the algebra $C(P)$ of complex functions $\{f,g,\cdots\}$ on the total 
space $P$ with typical points $\{p,q,\cdots\}$. Multiplication in $\Total$ is pointwise and the unit\footnote{The identity mapping on a set ${\cal S}$ is denoted by $\Id_{\cal S}$.} will be denoted by $\een_\Total$.
The algebra of complex functions $\{\alpha,\beta,\cdots\}$ on the (compact) structure group $G$,with typical elements $\{a,b,\cdots\}$ and neutral $e$, forms a Hopf algebra 
$\Hopf=C(G)$, with again pointwise multiplication, unit $\een_\Hopf$ and the coproduct and co-unit given by $[\Delta(\alpha)](a,b)=\alpha(ab)$ and $\epsilon(\alpha)=\alpha(e)$.
The co-action of $\Hopf$ on $\Total$ is given by the right action of $G$ on $P$, $(p,a)\rightarrow p\triangleleft a$. Identification of $C(P)\otimes C(G)$ with $C(P\times G)$ yields
\be\label{klassiek1}
\Delta_R:\Total\rightarrow\Total\otimes\Hopf:f(p)\rightarrow(\Delta_Rf)(p,a)=f(p\triangleleft a)\,,\ee
which obeys the axioms\footnote{The twist mapping $\sigma_{12}$ interchanges the factors of a tensor product: $\sigma_{12}(u\otimes v)=v\otimes u$.} (\ref{totaal1}) and  (\ref{totaal2}) below.
\[(\Delta_R\otimes\Id_\Hopf)\circ\Delta_R=(\Id_\Total\otimes\Delta)\circ\Delta_R\;;\;
f((p\triangleleft a)\triangleleft b)=f(p\triangleleft(ab))\;;\]
\be\label{totaal1}
(\Id_\Total\otimes\epsilon)\circ\Delta_R=\Id_\Total\;;\;f(p\triangleleft e)=f(p)\;.\ee
\[\Delta_R\circ m_\Total=(m_\Total\otimes m)\circ
(\Id_\Total\otimes\sigma_{12}\otimes\Id_\Hopf)\circ
(\Delta_R\otimes\Delta_R)\;;\;(f\cdot g)(p\triangleleft a)=f(p\triangleleft a)\,g(p\triangleleft a)\,;\]
\be\label{totaal2}
\Delta_R\circ\imath_\Total=\imath_\Total\otimes\imath\;;\;
\een_\Total(p\triangleleft a)=\een_\Total(p)\,\een_\Hopf(a)\;.
\ee
The freeness of the group action means that $(p\triangleleft a=p)\Leftrightarrow (a=e)$, and 
so the mapping $P\times G\rightarrow P\times P:(p,a)\rightarrow(p,p\triangleleft a)$ is injective. 
This freeness requirement leads to postulate the surjectivity of the dual map $\Total\otimes\Total\rightarrow\Total\otimes\Hopf$:
\be\label{vrij}
\Delta_{R1}=(m_\Total\otimes\Id_\Hopf)\circ(\Id_\Total\otimes\Delta_R)
:F(p,p^\prime)\rightarrow(\Delta_{R1}F)(p,a)=F(p,p\triangleleft a)\,.\ee
The invariant subalgebra of $\Total$ is defined as 
\be\label{base1}
\Total^\Hopf=\{f\in\Total\,|\,\Delta_Rf=f\otimes\een_\Hopf\,|\,f(p\triangleleft a)=f(p)\}\,.\ee
Thus a function $f(p)$ of $\Total^\Hopf$ depends only on the equivalence class of $p$ under the group action. These equivalence classes $x=[p]_G$ form the base space $B$ and there is a projection $\pi:P\rightarrow B:p\rightarrow x=\pi(p)$.
It is convenient to assume the existence of a {\it base algebra} $\Base=C(B)$ of functions $\{\widehat{f},\widehat{g},\cdots\}$, with an injective algebra homomorphism 
\be\label{base2}
\jmath:\Base\rightarrow\Total:\widehat{f}(x)\rightarrow (\jmath\widehat{f})(p)=\widehat{f}(\pi(p))\;;\;\jmath(\Base)=\Total^\Hopf\,.\ee
This mapping $\jmath$ is dual to the projection $\pi$ and the fact that $\jmath(\Base)$ should be the whole of $\Total^\Hopf$
amounts to the transitivity of the group action on each fibre.\\
A {\it topological quantum principal bundle} is defined as a quintuple $\{\Total,\Hopf,\Base,\Delta_R,\jmath\}$, where\\ {\bf 1)} $\{\Total,\Delta_R\}$ is a right comodule algebra under the Hopf algebra $\Hopf$, obeying (\ref{totaal1},\ref{totaal2}),\\ 
{\bf 2)} $\jmath:\Base\rightarrow\Total$ is an injective homomorphism such that (\ref{base2}) is satisfied and \\
{\bf 3)} $\Delta_{R1}$ in (\ref{vrij}) is surjective.\\
The {\it product bundle} is defined by:\\
{\bf 1)}
$\Total_\times=\Base\otimes\Hopf$ with $f\in\Total_\times$ given by $f(x,a)$ 
and $(\Delta_{R\times}f)(x,a;b)=f(x,ab)$.\\
{\bf 2)}
$[\jmath_\times \widehat{f}](x,a)=\widehat{f}(x)$.\\
{\bf 3)}
$\Delta_{R1\times}:\Total_\times\otimes\Total_\times\rightarrow\Total_\times\otimes\Hopf:
F(x,a;y,b)\rightarrow [\Delta_{R\times}F](x,a;b)=F(x,a;x,ab)$.\\
Obviously a surjective map, since $\forall\,H(x,a;b)\in\Total_\times\otimes\Hopf$, 
$F(x,a;y,b)\doteq H(x,a;a^{-1}b)$ transforms under  $\Delta_{R1\times}$ into $H(x,a;b)$. For these product bundles, conditions (\ref{vrij}) and (\ref{base2}) are automatically satisfied and do not need to be postulated separately.\\
A homo(iso)morphism of topological quantum principal bundles, over the same base algebra $\Base$ and with the same Hopf algebra $\Hopf$,
$\Psi :\{\Total^\prime,\Base,
\Hopf,\jmath^\prime,\Delta_R^\prime\}\rightarrow\{\Total,\Base,\Hopf,\jmath,\Delta_R\}$, is an algebra homo(iso)morphism $\Psi:\Total^\prime\rightarrow\Total$ such that 
$\jmath\circ\Psi=\jmath^\prime\;;\;\Delta_R\circ\Psi=
(\Psi\otimes\Id_\Hopf)\circ\Delta_R^\prime\,.$\\
A {\it trivial quantum principal bundle} is a topological quantum principal bundle 
$\{\Total,\Base,\Hopf,\jmath,\Delta_R\}$ with a {\it trivialisation}, i.e. an isomorphism 
with the product bundle 
$\{\Total_\times,\Base,\Hopf,\jmath_\times,\Delta_{R\times}\}$ :
$\Psi:\Base\otimes\Hopf\rightarrow\Total$.
Conditions above imply that $\Psi(\widehat{f}\otimes\een_\Hopf)=\jmath(\widehat{f})$ and 
$\Hopf$ co-acts as: $\Delta_R(\Psi(\widehat{f}\otimes\alpha))=\sum\Psi(\widehat{f}\otimes\alpha_1)\otimes\alpha_2$.
Since the requirements (\ref{vrij}) and (\ref{base2}) are satisfied in a product bundle, they will also be for a trivial bundle without imposing additional axioms. 
The homomorphism\footnote{In \cite{Severino,Hajac} $\Phi:\Hopf\rightarrow\Total$ is assumed to be only a linear (not an algebra) isomorphism and so is $\Psi$.} $\Phi:\Hopf\rightarrow\Total:
\alpha\rightarrow\Phi(\alpha)\doteq\Psi(\een_\Base\otimes\alpha)$ is dual to the mapping $\varphi:P\rightarrow G:p\rightarrow\varphi(p)$, satisfying $\varphi(p\triangleleft a)=\varphi(p)\,a$. Explicitely $(\Phi(\alpha))(p)=\alpha(\varphi(p))$. 
The map $\Psi:\Base\otimes\Hopf\rightarrow\Total$ is dual to  
$\psi:P\rightarrow B\times G:p\rightarrow\psi(p)=(\pi(p),\varphi(p))$. The action of $\Psi$ on $\Base\otimes\Hopf$ is then given by 
\be\label{isomorfisme}
\Psi:
\widehat{f}(x)\,\alpha(a)\rightarrow
\left(\Psi(\widehat{f}\otimes\alpha)\right)(p)
=\widehat{f}(\pi(p))\,\alpha(\varphi(p))\,.\ee
The co-action of $\Hopf$ on $\Phi(\alpha)$ is 
\be\label{coaktie1}
\Phi(\alpha)(p)\rightarrow
\left(\Delta_R(\Phi(\alpha))\right)(p;a)=\alpha(\varphi(p)\,a)\,.\ee
The inverse of $\Psi$ is 
\be\label{isoinvers}
\Psi^{-1}:f(p)\rightarrow 
(\Psi^{-1}f)(x,a)=f\left((p\triangleleft \varphi^{-1}(p))\triangleleft a\right)\;,\ee
well defined $\forall p\in\pi^{-1}(x)$, since 
$(p\triangleleft \varphi^{-1}(p))$ is invariant under the right group action.\\
The linear mappings of 
$HOM_{LIN}(\Hopf,\Total)$ form a convolution algebra 
with product and convolution unit given by
\be\label{convolutie}
(\Upsilon\star\Upsilon^\prime)(\alpha)\doteq
\sum\Upsilon(\alpha_1)\,\Upsilon^\prime(\alpha_2)\;;\;
{\bf 1}_\star(\alpha)\doteq{\bf 1}_\Total\,\epsilon(\alpha)\,.\ee
Since $\Phi$ is not only linear but also an algebra homomorphism it has a 
{\it convolution inverse}, which is an algebra anti-homomorphism given by: 
\be\label{inverse}
\left(\Phi^{[-1]}(\alpha)\right)(p)\doteq\alpha((\varphi(p))^{-1})\,.\ee
The co-action of $\Hopf$ on $\Phi^{[-1]}(\alpha)$ is 
\be\label{coaktie2}
\Phi^{[-1]}(\alpha)(p)\rightarrow
\left(\Delta_R(\Phi^{[-1]}(\alpha))\right)(p;a)=\alpha(a^{-1}\,(\varphi(p))^{-1})\,.\ee
Linear homomorphisms $\Upsilon\in Hom_{LIN}(\Hopf,\Total)$, are assumed to have a {\it spectral representation} of the form 
\be\label{spectraalvoorstelling}
(\Upsilon(\alpha))(p)=\int_G da\,\upsilon(p;a)\,\alpha(a)\,,\ee
If there is a bi-invariant Dirac measure $\delta(a,b)$ on $G$, the convolution product is  written as:
\[((\Upsilon\star\Upsilon^\prime)(\alpha))(p)=\int_G da\,db\,dc\,\upsilon(p;a)\upsilon^\prime(p;b)\delta(ab,c)\,\alpha(c)\;.\]
In particular, $\Phi$ is written as 
$(\Phi(\alpha))(p)=\int_G da\,\phi(p;a)\,\alpha(a)$. But, since it is an algebra homomorphism, one requires that $\phi(p;a)\phi(p;b)=\phi(p;a)\delta(a;b)$, which, under reasonable 
assumptions, means that $\phi(p;a)=\delta(\varphi(p);a)$.\\ 
A {\it gauge transformation} is a change of trivialisations: 
$\{\Psi:\Base\otimes\Hopf\rightarrow\Total\}\Rightarrow
\{\Psi^\prime:\Base\otimes\Hopf\rightarrow\Total\}$ and implements an automorphism of the product bundle 
$\Xi=\Psi^{-1}\circ\Psi^{\prime}:\Base\otimes\Hopf\rightarrow\Base\otimes\Hopf$.
Explicitely, it is given by:
\be\label{automorfisme1}
\Xi:\widehat{f}(x)\alpha(a)\rightarrow\left(\Xi(\widehat{f}\otimes\alpha)\right)(x,a)
=\widehat{f}(x)\,\alpha\left(\phi^\prime(p)\phi^{-1}(p)a\right)\,.\ee
In general, define the mapping:
\be\label{ijkverandering}
\tau:\Hopf\rightarrow\Base:\alpha\rightarrow\tau(\alpha)=
(\Id_\Base\otimes\epsilon)\left(\Xi(\een_\Base\otimes\alpha)\right)=\jmath^{-1}\left((\Phi^{\prime}\star\Phi^{[-1]})(\alpha)\right)
\,.\ee
Since $\widehat{f}\otimes\alpha=(\widehat{f}\otimes\een_\Hopf)(\een_\Base\otimes\alpha)=
(\een_\Base\otimes\alpha)(\widehat{f}\otimes\een_\Hopf)$, it can be shown that  $\widehat{f}\,\tau(\alpha)=\tau(\alpha)\,\widehat{f}$ for any $\widehat{f}$ which means that 
$\tau$ takes values in the center ${\bf Z}(\Base)$ of the algebra $\Base$. In our particular case, $\Base$ is commutative\footnote{this implies that ${\bf Z}(\Base)=\Base$.} and we have $\tau(\alpha)(x)=\alpha\left(\phi^\prime(p)\phi^{-1}(p)\right)$.
Conversely an algebra homomorphism 
$\tau:\Hopf\rightarrow{\bf Z}(\Base):\alpha(a)\rightarrow\tau(\alpha)(x)=\alpha(\widehat{\tau}(x))$, where 
$\widehat{\tau}:B\rightarrow G$, defines a bundle automorphism $^\tau\Xi$ of the product bundle $\Base\otimes\Hopf$. It is given by: 
\be\label{automorfisme2}
\left(^\tau\Xi(\widehat{f}\otimes\alpha)\right)(x,a)=
\widehat{f}(x)\,\alpha(\widehat{\tau}(x)a)\;.\ee
The homomorphisms $\tau\in HOM(\Hopf,{\bf Z}(\Base))$ form a group with the convolution product $\tau_1\star\tau_2$ as multiplication, with inverse $\tau^{[-1]}=\tau\circ S$ and neutral $\nu=
\een_\Hopf\,\epsilon$. It is isomorphic to the group of automorphisms of the product bundle. Indeed, from (\ref{automorfisme2}) it is straightforward to verify that 
$\quad^{(\tau_1\star\tau_2)}\Xi={^{\tau_1}\Xi}\,\circ{^{\tau_2}\Xi}\;;\quad
{^{(\tau\circ S)}\Xi}=({^{\tau}\Xi})^{-1}$.
\section{Differential principal fibre bundles}
\setcounter{equation}{0}
In the classical theory of Lie group bundles over differential manifolds, the infinitesimal group action allows to define the vertical subspace which is isomorphic to the group manifold.
Dual to the vertical vectors, one defines horizontal one-forms given a differential structure on the algebras $\Total$ and $\Base$. Any graded differential calculus over an algebra can be obtained from the {\it universal differential envelope} taking its quotient by a graded differential ideal. Compatibility conditions of the comodule algebra structure of $\Total$ and the differential calculus are then required. it happens that the most economic set of  definitions is obtained using the universal calculus. 
This universal calculus $\{\Omega^\bullet(\Total)=\bigoplus^\infty_0\,\Omega^{n}(\Total),\dif\}$ is realised as a subalgebra of the tensor algebra $\Omega^n(\Total)\subset\bigotimes^{n+1}\Total$.
When $\Total=C(P)$ and $\bigotimes^{n+1}\Total$ is identified as functions of $(n+1)$ points of $P$, $\Omega^n(\Total)$ is given by those functions, vanishing on adjacent diagonals:
$\{F_n(p_0,p_1,\cdots,p_n)\;|\;p_j=p_{j+1}\Rightarrow F_n(\cdots)=0\}$,
The product is defined by concatenation:
$\{F_n\cdot G_m\}(p_0,p_1,\cdots,p_n,p_{n+1},\cdots,p_{n+m})=
f_N(p_0,p_1,\cdots,p_n)\,G_m(p_n,p_{n+1},\cdots,p_{n+m})$. \\
The differential is 
$\{\dif F_n\}(p_0,p_1,\cdots,p_n,p_{n+1})=\sum_{j=0}^{n+1}\,
(-1)^{j}\,F_n(p_0,p_1,\cdots,\widehat{p_j},\cdots,p_{n+1})$ and obeys the Leibniz rule.
Obviously $\Omega^\bullet(\Total)$ is a $\Total$-bimodule. On $\Base$ there is a similar calculus 
$\{\Omega^\bullet(\Base)$ such that the injective homomorphism $\jmath:\Base\rightarrow\Total$ extends to the universal envelope:
$\jmath:\Omega^{n}(\Base)\rightarrow\Omega^{n}(\Total)$.
The space of {\it horizontal one-forms}, $\Gamma^1_{hor}(\Total)=\Total\Omega^{1}(\Base)\Total$, consists of linear combinations of terms of the form  $f(p)\widehat{F}_1(\pi(p),\pi(p^\prime))g(p^\prime)$ or even $f(p)[\widehat{h}(\pi(p^\prime)-\widehat{h}(\pi(p))]g(p^\prime)$, where $f$ and $g$ are in $\Total$, $\widehat{F}_1$ in $\Omega^1(\Base)$ and $\widehat{h}$ belongs to $\Base$. If the map $\Delta_{R1}$ of (\ref{vrij}) is to be surjective, so must be its restriction to $\Omega^{1}(\Total)\subset\Total\otimes\Total$, denoted by the same symbol $\Delta_{R1}$.
\be\label{vrijbis}
\Delta_{R1}: \Omega^{1}(\Total)\rightarrow\Total\otimes Ker\{\epsilon\}:
F_1(p,p^\prime)\rightarrow(\Delta_{R1}F_1)(p;a)=F_1(p,p\triangleleft a)
\,.\ee
Since $(\Delta_{R1}F_1)(p;e)=F_1(p,p)=0$, its image lies in $\Total\otimes Ker\{\epsilon\}$\footnote{The kernel $Ker\{\epsilon\}$ is an ideal of $\Hopf$ which may be decomposed as $\Hopf=Ker\{\epsilon\}\oplus{\bf C}\een_\Hopf$. The projection $\pi_\epsilon$ on $Ker\{\epsilon\}$ is given by $\widetilde{\alpha}\doteq\pi_\epsilon(\alpha)=\alpha-\epsilon(\alpha)\een_\Hopf$.}.
Now, any $F_{1;hor}\in\Gamma^1_{hor}(\Total)$ satisfies 
$\Delta_{R1}F_{1;hor}=0$ and so $\Gamma^1_{hor}(\Total)\subset Ker(\Delta_{R1})$, but the converse $\Gamma^1_{hor}(\Total)\supset Ker(\Delta_{R1})$ does not necessarely holds. When $\Delta_{R1}$ is surjective and $\Gamma^1_{hor}(\Total)\equiv Ker(\Delta_{R1})$, an exact sequence is obtained:
\be\label{exakterij}
0\rightarrow\Gamma^1_{hor}\hookrightarrow
\Omega^1(\Total)\stackrel{\Delta_{R1}}{\rightarrow}
\Total\otimes Ker\{\epsilon\}\rightarrow 0\;.\ee
The sextuple  
$\{\Total,\Base,\Hopf,\jmath,\Delta_R,\Omega^\bullet(\Total)\}$ is then called a {\it differential quantum principal fibre bundle} (with universal differential calculus!).\\
When there is a trivialisation $\Psi:\Base\otimes\Hopf\Rightarrow\Total:f_\times(x,a)\rightarrow
(\Psi f_\times)(p)=f_\times(\pi(p),\phi(p))$, it can be shown that 
the above sequence (\ref{exakterij}) is always exact: 
\section{Connection and Curvature}
\setcounter{equation}{0}
A {\it connection} is a splitting of the sequence (\ref{exakterij}) by a left $\Total$-module homomorphism $\sigma:\Total\otimes Ker\{\epsilon\}\rightarrow \Omega^1(\Total): F(p,a)\rightarrow(\sigma F)(p,p^\prime)$, i.e. such that 
\be\label{spleten}
\Delta_{R1}\circ\sigma=\Id_{\Total\otimes Ker\{\epsilon\}}\;;\;
(\Delta_{R1}(\sigma F))(p,a)=(\sigma F)(p,p\triangleleft a)=F(p,a)\,.\ee
This defines the vertical projection 
\be\label{vertikaleprojektie}
\Pi_{ver}=\sigma\circ\Delta_{R1}:\Omega^1(\Total)\rightarrow\Omega^1(\Total):f\dif g\rightarrow
fg_1\,\sigma(\een_\Total\otimes\pi_\epsilon(\beta_2))\,.\ee
Its image is called the {\it subspace of vertical one-forms} $\Gamma^1_{ver}(\Total)$ whose direct sum with $\Gamma^1_{hor}(\Total)$ results in $\Omega^1(\Total)=\Gamma^1_{hor}(\Total)\oplus\Gamma^1_{ver}(\Total)$.
In diagrammatic form :
\[\begin{array}{ccccccccc}0&\rightarrow&\Gamma^1_{hor}(\Total)&\hookrightarrow&
\Omega^1(\Total)&\stackrel{\Delta_{R1}}{\longrightarrow}&\Total\otimes Ker\{\epsilon\}&
\rightarrow&0\\
&&&&\downarrow&&\downarrow&&\\
&&&&\Pi_{ver}\downarrow&\stackrel{\sigma}{\swarrow}&\downarrow{\bf id}&&\\
&&&&\downarrow&&\downarrow&&\\
&&&&\Omega^1(\Total)&\stackrel{\Delta_{R1}}{\longrightarrow}&\Total\otimes Ker\{\epsilon\}&&
\end{array}\]
Furthermore this splitting should be right covariant in the sense below. The {\it right adjoint coaction} of $\Hopf$ on itself is given by : 
${\bf Ad}_R:\Hopf\rightarrow\Hopf\otimes\Hopf:\alpha(a)\rightarrow({\bf Ad}_R\alpha)(a,b)=
\alpha(b^{-1}ab)$.
When restricted to $\widetilde{\alpha}\in Ker\{\epsilon\}$ its image lies in $Ker\{\epsilon\}\otimes\Hopf$ and so $Ker\{\epsilon\}$ is a $\Hopf$-comodule. The tensor product of the $\Hopf$-comodules $\Total$ and $Ker\{\epsilon\}$ is a $\Hopf$-comodule with coaction 
\be\label{Deltatilde}
\widetilde{\Delta_R}:\Total\otimes Ker\{\epsilon\}\rightarrow
\Total\otimes Ker\{\epsilon\}\otimes\Hopf:
A(p,a)\rightarrow(\widetilde{\Delta_R}A)(p,a,b)=A(p\triangleleft b,b^{-1}ab)\,.\ee
The tensor product $\Total\otimes\Total$ also is a right $\Hopf$-comodule.
It remains a right $\Hopf$-comodule when restricted to $\Omega^1(\Total)$ with coaction denoted by $\Delta_R^{\prime}$:
\be\label{Deltapriem}
\Delta_{R}^{\prime}:
\Omega^1(\Total)\rightarrow\Omega^1(\Total)\otimes\Hopf: F_1(p,p^\prime)\rightarrow
(\Delta_R^\prime F_1)(p,p^\prime,a)=F_1(p\triangleleft a,p^\prime\triangleleft a)\,.\ee
Right covariance requires $\sigma$ to intertwine these coactions:
$(\sigma\otimes{\bf id}_\Hopf)\circ\widetilde{\Delta_R}=
\Delta_R^{\prime}\circ\sigma$.\\
Since $\sigma$ is an homomorphism of left $\Total$-moduli, it is defined by its action on ${\bf 1}_\Total\otimes Ker\{\epsilon\}$. The {\it connection 1-form} is the linear map: 
$\Theta_{Ker}:Ker\{\epsilon\}\rightarrow\Omega^1(\Total):
\widetilde{\alpha}\rightarrow\sigma(\een_\Total\otimes\widetilde{\alpha})\,$,
which obeys the splitting property (\ref{spleten}): $\Delta_{R1}\circ\Theta_{Ker}={\bf 1}_\Total\otimes\Id_{Ker\{\epsilon\}}$ and the right covariance : 
$\Delta_R^{\prime}\circ\Theta_{Ker}=(\Theta_{Ker}\otimes\Id_\Hopf)\circ{\bf Ad}_R$.\\
Since $\Hopf=Ker\{\epsilon\}\oplus{\bf C}{\bf 1}_\Hopf$, the map 
$\Theta_{Ker}$ can be extended to the whole of $\Hopf$ by 
\be\label{volle}
\Theta:\Hopf\rightarrow\Omega^1(\Total):\alpha\rightarrow
\Theta(\alpha)=\Theta_{Ker}(\pi_\epsilon(\alpha))=\Theta_{Ker}(\alpha-\epsilon(\alpha){\bf 1}_\Hopf)\,,\ee
which is also called the {\it connection 1-form} and obeys:
\be\label{connectievoorwaarden}
\Theta({\bf 1}_\Hopf)=0\;,\;
\Delta_{R1}\circ\Theta={\bf 1}_\Total\otimes\pi_\epsilon\;,\;
\Delta_R^{\prime}\circ\Theta=(\Theta\otimes\Id_\Hopf)\circ{\bf Ad}_R\,.\ee
A spectral representation is assumed:  
$(\Theta(\alpha))(p,p^\prime)=\int_Gdc\,\theta(p,p^\prime;c)\,\alpha(c)$, such that 
(\ref{connectievoorwaarden}) can be written as conditions on the {\it spectral density} $\theta(p,p^\prime;c)$.\\
In a trivial bundle, there always exists a {\it trivial connection}:
\be\label{trivialeconnectie} 
\Theta^\Phi(\alpha)=(\Phi^{[-1]}\star\dif\Phi)(\alpha)=\sum\Phi^{[-1]}(\alpha_1)\dif\Phi(\alpha_2)
\;,\ee
with spectral density 
$\theta^\Phi(p,p^\prime;a)=\delta(\varphi^{-1}(p)\varphi(p^\prime);a)-\delta(e;a)$\\
A generic connection can be written in the form:
\be\label{connectiebis}
\Theta=\Phi^{[-1]}\star\dif\Phi+\Phi^{[-1]}\star\Gamma\star\Phi\,.\ee
With $(\Gamma\alpha)(p,p^\prime)
=\int_G\,da\,\gamma(p,p^\prime;a)\,\alpha(a)$, the spectral density $\gamma$ obeys:
\[\int_Gdc\,\gamma(p,p^\prime;c)=0\;,\;
\gamma(p,p\triangleleft a;c)=0\;,\;
\gamma(p\triangleleft a,p^\prime\triangleleft a;c)=
\gamma(p,p^\prime;c)\,.\]
Following \cite{Hajac}, a connection is said to be a {\it strong connection} if 
$(\Gamma(\alpha))(p,p^\prime)$ is a {\it strongly horizontal 1-form} of $\Omega^1_{shor}(\Total)\doteq\jmath(\Omega^1(\Base))$\footnote{More generally, the ${\cal P}$-bimodule of {\it horizontal n-forms} is defined by: $\Gamma^n_{hor}=\{\Gamma^1_{hor}\cdots\Gamma^1_{hor}\}$ with n factors, or also ${\cal P}\{\jmath((\Omega^1(\Base)){\cal P})\cdots(\jmath(\Omega^1(\Base))\}{\cal P})$
and the left ${\cal P}$-module of {\it strongly horizontal} n-forms, by: 
$\Gamma^n_{shor}=\jmath(\Omega^n(\Base)){\cal P}$.}. In this case, $\gamma(p,p^\prime;a)=\widehat{\gamma}(\pi(p),\pi(p^\prime);a)$, 
where $\widehat{\gamma}(x,x^\prime;a)$ 
obeys $\widehat{\gamma}(x,x;a)=0\,;\;\int_Gda\,\widehat{\gamma}(x,x^\prime;a)=0$.
For these strong connections:
\be\label{sterkebis}
(\Theta\alpha)(p,p^\prime)=
\int_Gda\,[\delta(e;a)+\widehat{\gamma}(x,x^\prime;a)]\,
\alpha(\varphi^{-1}(p)\,a\,\varphi(p^\prime))\;-\;\alpha(e)\,.\ee
A connection will be called a {\it classical connection} if, besides this, there exists a map $\widehat{g}:B\times B\rightarrow G:(x,x^\prime)\rightarrow \widehat{g}(x,x^\prime)$ such that 
$\delta(e;a)+\widehat{\gamma}(x,x^\prime;a)=\delta(\widehat{g}(x,x^\prime);a)$, 
and $\widehat{g}(x,x)=e$. These classical connections are written as:
\be\label{klassiekeconnectie}
(\Theta(\alpha))(p,p^\prime)=\alpha(a^{-1}\,\widehat{g}(x,x^\prime)\,a^\prime)-\alpha(e)\,,\ee
where $(x,a)=(\pi(p),\phi(p))$ and $(x^\prime,a^\prime)=(\pi(p^\prime),\phi(p^\prime))$.\\
With the same $\Gamma:\Hopf\rightarrow\Omega^1(\Base)$ and another 
trivialisation, the connection  (\ref{connectiebis}) 
may be written as $
^\tau\Theta=(^\tau\Phi)^{[-1]}\star{\bf d}\,(^\tau\Phi)+
(^\tau\Phi)^{[-1]}\star\Gamma\star\,(^\tau\Phi)= 
\Phi^{[-1]}\star{\bf d}\Phi+
\Phi^{[-1]}\star\,^\tau\Gamma\star\Phi$, where 
$^\tau\Gamma=\tau^{[-1]}\star{\bf d}\tau+
\tau^{[-1]}\star\Gamma\star\tau$ defines a new connection $^\tau\Theta$ in the original trivialisation\footnote{This is the so-called {\it active viewpoint} under which $\tau\in{\cal G}$ acts on the space of connections.}, {\it gauge equivalent} with the original. 
The spectral density of the transformed connection reads:
\be\label{ijkveranderdedichtheid}
^\tau\widehat{\gamma}(x,x^\prime;a)=
\delta(e;\widehat{\tau}(x)a\widehat{\tau}^{-1}(x^\prime))-\delta(e;a)+ \,\widehat{\gamma}(x,x^\prime;\widehat{\tau}(x)a\widehat{\tau}^{-1}(x^\prime))
\,.\ee
For a classical connection (\ref{klassiekeconnectie}) the spectral density is given as  
\be\label{klassiekeijkveranderdegroep}
\delta(e;a)+{^\tau\widehat{\gamma}}(x,x^\prime;a)=
\delta(^\tau\widehat{g}(x,x^\prime);a)\;;\;
{^\tau\widehat{g}}(x,x^\prime)=
\widehat{\tau}^{\,-1}(x)\,\widehat{g}(x,x^\prime)\,\widehat{\tau}(x^\prime)\;.\ee
As in the classical case, one defines the covariant differential {\bf D} of horizontal 1-forms and its square ${\bf D}^2$ leads to define the curvature as a mapping ${\bf F}:{\cal H}\rightarrow\Omega^2({\cal P})$
\be\label{kromming2}
{\bf F}=d\Theta+
\Theta\star_\Delta\Theta\nonumber=
\Phi^{[-1]}\star\left(\dif\Gamma+\Gamma\star\Gamma\right)\star\Phi
\,.\ee
For classical connections, with $p=\psi^{-1}(x,a)$ etc., the curvature in (\ref{kromming2}) is obtained as  
\be\label{kromming4}
({\bf F}\alpha)(p,p^\prime,p^{\prime\prime})=
\alpha\Bigl(a^{-1}\widehat{g}(x,x^\prime)\widehat{g}(x^\prime,x^{\prime\prime})a^{\prime\prime}
\Bigr)-\alpha\Bigl(
a^{-1}\widehat{g}(x,x^{\prime\prime})a^{\prime\prime}\Bigr)\,.
\ee
A locally trivial quantum principal bundle \cite{BudKon,Calow} is defined by a covering of the algebra $\Total$ which encodes the topology of the quantum bundle. This is explained in detail in the Poster session P112 for the Dirac monopole on the two-sphere.
%

\end{document}